\documentclass[11pt,twoside]{article}
 
\usepackage{asp2004}
\usepackage{graphicx}
\usepackage{epsf}
\usepackage{epsfig}
\usepackage{lscape}
\usepackage{amssymb}
 
\begin{document}

\title{Tiny-Scale Atomic Structure and the Cold Neutral Medium---Review and Recap}   
\author{Carl Heiles}   
\affil{Astronomy Department, University of California, Berkeley CA}    
 
\begin{abstract} Almost a decade ago I wrote an article with the same
  title as this. It focused on the physical properties of the Tiny-Scale
  Atomic Structure (TSAS) as discrete structures of the Cold Neutral
  Medium (CNM). To be observable, tiny discrete structures that don't
  grossly violate pressure equilibrium need two attributes: low
  temperatures and geometrical anisotropy.  Here I update that
  article. I discuss thermal and pressure equilibrium, ionization,
  optical lines, H$_2$ abundance, and evaporation. \end{abstract}


\section{Canonical Values for TSAS as Discrete Structures} \label{one}

Following Heiles (1997; H97), we use currently-available observations of
the tiny-scale atomic structure (TSAS) to describe its properties; the
observations have improved, so the inferred properties are now somewhat
different. The TSAS is seen in 21-cm line absorption so must be part of
the Cold Neutral Medium (CNM). The observations consist of time-variable
21-cm absorption lines against pulsars, VLBI images of the 21-cm line in
absorption against extended radio sources, and optical absorption line
observations against background stars. The canonical values for
important properties come directly from these observations and have no
theoretical or otherwise imaginary input, except for the subjective
judgment and weighting that we must perform to collapse a variety of
measured values into a single number that we can use as a canonical
value. We provide values for the following properties: \begin{enumerate}

\item TSAS column density:

\begin{equation} \label{tsascoln}
N(HI)_{TSAS, 18} = 0.75 \, T_{15} \ , 
\end{equation}

\noindent where $N(HI)_{18}$ is the HI column density in units of
$10^{18}$ cm$^{-2}$ and $T_{15}$ the temperature in units of 15 Kelvins
(\S \ref{equiltemp} justifies this canonical value of 15 K). The
$T$-dependence arises because the 21-cm line opacity $\propto {1 \over
T}$. The numerical value 0.75 is 4 times larger than the canonical
value adopted by H97 and is based on Weisberg \& Stanimirovic (this
meeting).

\item The TSAS scale length in the plane of the sky (denoted by the
  subscript $\perp$):

\begin{equation} \label{tsaslperp}
L_{TSAS, \perp} = 30 \ {\rm AU} \ .
\end{equation}

\noindent The $\perp$ arises because with time variability and angular
structure, it's the plane-of-the-sky length that's relevant. 

\item The line-of-sight scale length $L_{TSAS,||}$ can be different (by a
factor $G$) because it is unlikely for the structures to be isotropic:

\begin{equation}
G \equiv { L_{TSAS,||} \over L_{TSAS,\perp}} \ .
\end{equation}

\item We combine $N(HI)_{TSAS}$ and $L_{TSAS,||}$ to obtain the typical TSAS
  volume density:

\begin{equation} \label{nvol}
n(HI)_{TSAS} \equiv {N(HI)_{TSAS} \over L_{TSAS,||}} = 1700 \ {T_{15}
  \over G} \ {\rm cm^{-3}} \ .
\end{equation}

\item The canonical values for volume density and temperature lead to
the pressure $\widetilde{P} \equiv {P \over k} = nT$:

\begin{equation} \label{peqn1}
\widetilde{P}_{TSAS} = 26000 \ {T_{15}^2 \over G} \ .
\end{equation}

\end{enumerate}

With the above we make a simple, highly important observation: equation
\ref{peqn1} shows that unless the TSAS is cold and/or anisotropic, its
pressure greatly exceeds not only the typical CNM thermal pressure of
$\widetilde{P}_{CNM} \sim 4000$ cm$^{-3}$ K, but also the Galactic
hydrostatic pressure of $\widetilde{P}_{4K} \sim 28000$ cm$^{-3}$ K
(Boulares \& Cox 1990). This single point dominates the rest of the
discussion, which includes some theoretical considerations.

\section{Thermal Equilibrium, Thermal Pressure, and Total Pressure}
\label{two}

Is the TSAS in thermal equilibrium?  We normally expect the CNM to
attain thermal equilibrium because the time scale is short.  We have
(Wolfire et al.\ 2003; WMHT)

\begin{equation}
t_{cool}= {5 \over 2} \ {1.1 \ nkT \over n^2 \Lambda} \ ,
\end{equation}

\noindent where the symbols have their usual meanings and $n^2 \Lambda$
is the cooling rate per cm$^{-3}$; cooling varies as $n^2$ because it is
a collisional process.  For collisional excitation of CII (the 158
$\mu$m line) by H atoms at $T=15$ K, which is the likely temperature of
the TSAS, we have

\begin{equation}
t_{cool} = 1.6 \times 10^4 \ G \ 460^{\left({1 \over T_{15}} - 1 \right)}
\ {\rm yr} \ .
\end{equation}

\noindent or

\begin{equation}
t_{cool} = 1.0 \times 10^5 \ {T_{15}^2 \over 
\widetilde{P}_{TSAS}/\widetilde{P}_{CNM}} \ 460^{\left({1 \over T_{15}} - 1 \right)}
\ {\rm yr} \ .
\end{equation}

\noindent Here $\widetilde{P}_{CNM}= 4000$ cm$^{-3}$ K, our adopted
pressure for the CNM. If the TSAS has ${T_{15}^2 \over
\widetilde{P}_{TSAS}/\widetilde{P}_{CNM}} \sim 1$, then $t_{cool} \sim
10^5$ yr. This is short by most interstellar standards.  However, for
the time- and space-variable TSAS it might not be short, so we
consider two cases, the thermal equilibrium case and the
nonequilibrium case.

\subsection{Thermal Equilibrium Case}

\begin{figure}[h!]
\includegraphics[scale=0.6]{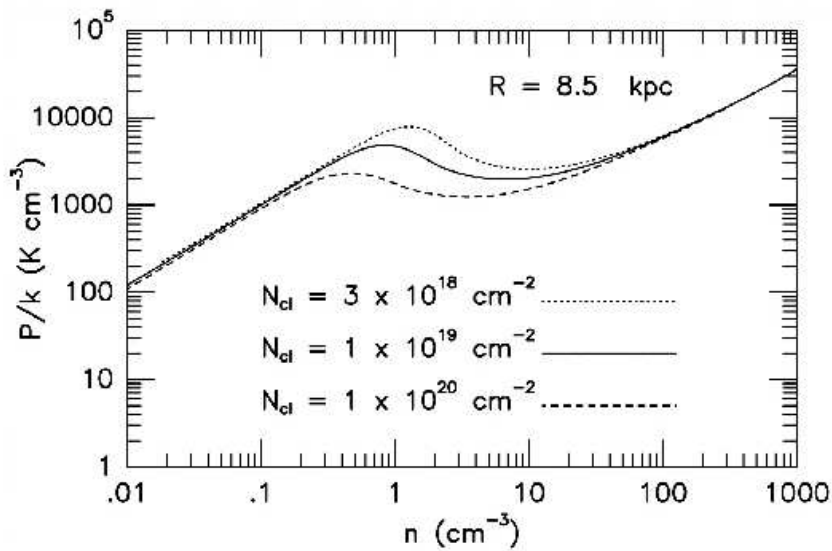}
\caption{The thermal-equilibrium pressure $\widehat{P}$ versus
  volume density $n(HI)$. Figure 9 from WMHT. \label{wolf9}}
\end{figure}

Suppose first that the TSAS is in thermal equilibrium.  Then the density
and pressure are related as shown in Figure \ref{wolf9}, which is Figure
9 from WMHT. For discussion purposes, we use the solid curve for
$N(HI)_{19}=1$ because we envision the TSAS to be embedded in CNM.  At
low and high $n(HI)$ it defines the Warm and Cold Neutral Media with
temperatures $T \sim 8310$ and $\sim 62$ K, respectively.

The CNM thermal pressure has a minimum ($\widetilde{P}_{CNM,min} \sim
1960$ cm$^{-3}$ K) located near $n= 10$ cm$^{-3}$. This minimum thermal
pressure is defined by microscopic heating and cooling processes, which
simply and unequivocally means that the CNM cannot exist in thermal
equilibrium at lower thermal pressures. This is a strict lower limit for
the CNM thermal pressure in thermal equilibrium. Similarly, the WNM has
a microphysical-process thermal pressure maximum
($\widetilde{P}_{WNM,max} \sim 4810$ cm$^{-3}$ K) located near $n= 1$
cm$^{-3}$.

There is no such strict microphysical-process upper limit on the CNM
thermal pressure. However, there is an upper limit that arises from
pressure-balance considerations.  The CNM occupies a small fraction of
the total interstellar volume, so it has to be embedded in warmer gas,
either the WNM or the Hot Ionized Medium (HIM). If the interface is not
dynamic it must have pressure balance.  If the bounding medium is the
WNM with no magnetic field, then the CNM thermal pressure must lie below
$\widetilde{P}_{WNM,max}$; otherwise, the WNM is forbidden. 

However, for dynamical equilibrium it's the {\it total} pressure that
counts, not just the thermal pressure. The WNM probably has the typical
interstellar magnetic field of $\sim 6$ $\mu$G, which exerts a magnetic
pressure of $\widetilde{P}_{mag} \lesssim 10^4$ cm$^{-3}$ K. This
$\lesssim$ sign is important: it's the magnetic pressure {\it
difference} that counts, and even then only for that component that is
perpendicular to the field lines, so if the field lines don't lie along
the interface, or if the CNM field is comparable to the WNM field, then
the magnetic pressure difference is less. There's a {\it strict} upper
limit to the CNM pressure if it's in thermal and dynamical equilibrium,
which is the hydrostatic pressure of the Galactic disk; this is about
$\widetilde{P} \sim 28000$ cm$^{-3}$ K (Boulares \& Cox 1990). If
there's no magnetic field and no turbulence, then the CNM pressure has
to lie in the very narrow range $\widetilde {P} \sim 1960$ to 4810
cm$^{-3}$ K, which is a factor of only 2.5.

In equation \ref{peqn1}, for the {\it strict} upper limit $\widetilde{P}
\sim 28000$ cm$^{-3}$ K, we require ${T_{15}^2 \over G} \lesssim 1$.
It's unlikely, in our opinion, that the strict upper limit applies
always. A more reasonable value is lower, perhaps $\widetilde{P} \sim
10000$ cm$^{-3}$ K, for which we require ${T_{15}^2 \over G} \sim
0.4$. Below we will find that it is impossible to attain $T \lesssim 12$ K, so
this implies $G \sim 2.5$. This seems like a perfectly reasonable degree
of anisotropy for TSAS, particularly in light of H97's estimate as $G
\sim 10$ for a reasonable choice if TSAS is sheetlike.

\subsection{Thermal Nonequilibrium Case}

	If we don't have thermal equilibrium all bets are off and we can
invoke any conditions we please.  From the standpoint of making TSAS
more easily explained by individual structures, we move toward smallness
by invoking higher densities and pressures.  For example, if $\widetilde
{P}_{TSAS} = 40000$ cm$^{-3}$ K, ten times higher than the equilibrium
$\widetilde{P}_{CNM}$, $n_{TSAS}$ increases and $t_{cool}$ decreases by
a factor of ten.  The decreased cooling time presents the quandary that
it's harder to escape thermal equilibrium!

Raising the temperature above the lowest equilibrium value of 12 K is
counterproductive in the sense of requiring higher HI column densities
to produce observable 21-cm line absorption. Raising temperature instead
of density also increases the cooling rate even more, because of the
highly nonlinear factor $e^{-92 {\rm K} \over T}$ in the cooling
function $\Lambda$ (equation \ref{ccool}; for example, if we increase
the temperature from 16 K to a typical CNM temperature of 60 K, the
cooling rate $\Lambda$ would increase by a factor 100 (!).

Conclusion: if we invoke nonequilibrium, it's easier to obtain the
observed TSAS by keeping it cold and increasing the volume
density. This, of course, also increases the pressure, which leads to
some spectacular dynamics\dots

\subsection{Dynamical Nonequilibrium Case} \label{dynamics}

In the absence of pressure equilibrium, tiny-scale structures would
expand rapidly into the surrounding medium. Soon the structures would no
longer be ``tiny''. For example, for a pressure ratio between TSAS and
the surrounding medium of 10 the expansion velocity is close to Mach 10,
or $\sim 0.3$ km s$^{-1}$; it takes only 500 years to expand by the TSAS
characteristic scale size of 30 AU. So a structure won't last long. On the
other hand, one of the ways we detect TSAS is by time
variability---which is a natural result of dynamical nonequilibrium!
So, in this sense, dynamical nonequilibrium is suggested by the
observations. 

If the TSAS pressure is high, then you have to answer the question: how
did the TSAS pressure get so high to begin with? The answer might be
intermittency in the turbulence of the enveloping medium (Hily-Brandt,
this meeting; Coles, this meeting).

	To summarize: The nonequilibrium case has intrinsically high
time variability.  In this sense, it satisfies one of the observational
signatures that define TSAS! This poses the question in a slightly
different way: the observed time variability shows that TSAS is not in
perfect dynamical equilibrium, but is it {\it far} from or {\it not too
far} from equilibrium?

\section{Physical Properties of the TSAS in Equilibrium} \label{three}

\subsection{The Equilibrium Temperature} \label{equiltemp}

The equilibrium temperature is determined by equality of heating and
cooling rates.  Regarding {\it cooling}: The TSAS has no observed
molecules, so we are left with the only effective atomic coolant, which
is collisional excitation of CII. For cold gas, Carbon is almost fully
ionized and Hydrogen almost fully neutral, so $n_e$ is almost equal to
the gaseous Carbon abundance. We assume that the C/H abundance is $4
\times 10^{-4}$ by number and that the {\it gaseous} C/H ratio is
$\delta_c$ times smaller (some C goes into dust; people typically assume
that the depletion factor $\delta_c = 0.35$).  We use WMHT equations C1
and C2 to write

\begin{equation} \label{ccool}
n^2 \Lambda = 7.1 \times 10^{-27} \delta_c \, n(HI)^2 \
\left[ T_{15}^{0.07} + \, 0.41 \, \delta_c \, T_{15}^{-0.5} \right] \, 
e^{-92/T} \ \rm{erg \ cm^{-3} \ s^{-1}} \ .
\end{equation}

\noindent Note that there are two submechanisms: collisions by atomic HI
(the first term) and by electrons (the second term). The H-collisions
dominate, even there are no grains and $\delta_c = 1$.

{\it Heating} is another matter.  The commonly-accepted dominant heating
mechanism is photoionization of PAH; this provides typical CNM
temperatures of $\sim 60$ K.  We envision TSAS temperatures to be much
smaller, so we assume specifically that PAH heating (and, by
implication, PAHs themselves) are absent.  This has the ancillary
consequence of freeing much of the Carbon to the gas phase, making
$\delta_c \rightarrow 1$.

Without PAH heating, we have two possibly important heating mechanisms:
photoionization of CI by starlight and ionization of HI by cosmic rays
and soft X-rays. We first consider the former mechanism alone, which
provides an absolute lower limit for the temperature of HI regions, and
then include the latter.

\subsubsection{ Heating Exclusively by Photoionization of CI by Starlight}

For heating by photoionization of CI, the heating rate per unit volume
$n \Gamma$ is proportional to the photoionization rate per C atom
multiplied by the volume density of C atoms $n(CI)$. In ionization
equilibrium, the photoionization rate is equal to the recombination rate,
and because Carbon is almost fully ionized the recombination rate is
much easier to calculate. This gives

\begin{equation}
n \Gamma = \underbrace{ n(CII) \, n_e \, \alpha(T)}_{\rm ioniz \ rate \ per \
  cm^{-3}} \times \ \langle E_i \rangle  \ ,
\end{equation}

\noindent where $\alpha$ is the recombination coefficient. $E_i$ is the
mean energy gained per photoionization; the UV radiation field drops to
zero above 13.6 eV, so $E_i \approx (13.6\; {\rm eV} - I\! P_C)/2$
(where $I \!P_C$ is Carbon's ionization potential in eV; see Spitzer
1978, p.\ 107, p.\ 134-136, p.\ 142-143). Remembering that nearly all
the electrons are contributed by gaseous Carbon, and that Carbon is
almost fully ionized, we have

\begin{equation} \label{cheat}
n \Gamma \approx 6.4 \times 10^{-30} \delta_c^2 \, n(HI)^2 \, T_{15}^{-0.5}
\end{equation}

We get the temperature by equating equations \ref{ccool} and
\ref{cheat}. If atomic H cooling (the first term in equation \ref{ccool})
is negligible, then (amazingly enough), as pointed out by Spitzer (1978,
p.\ 143), we are left with the ultra-simple equation

\begin{equation} 
e^{-92/T}= 0.0022 \ {\rm , \  or} 
\end{equation}
$$T = 15.03 \ {\rm K} \ . $$

\noindent This is a very robust result: it is independent of both the
volume density (as long as $n(HI) \lesssim 3000$ cm$^{-3}$, where
collisional de-excitation of CII becomes important) and the starlight
intensity! Now, as we remarked above, H cooling does in fact dominate,
which lowers the equilibrium temperature to about 12 K. This is a lower
limit for the temperature in purely atomic regions.

\subsubsection{ The Heating Contribution from HI Ionization}

The neural ISM has no UV photons above 13.6 eV energy, so HI can be
ionized only by cosmic rays and soft X-rays. Soft X-rays are attenuated
by small columns of HI, so their heating rate is sensitive to
shielding. WMHT discuss this rather deeply and conclude that a typical
primary ionization rate from soft X-rays is $\zeta_X \sim 1.6 \times
10^{-17}$ s$^{-1}$ per H atom. 

The primary ionization rate from cosmic rays $\zeta_{CR}$ is highly
uncertain.  For example, WMHT adopt $\zeta_{CR,WMHT} = 1.8 \times
10^{-17}$ sec$^{-1}$ per H-atom, a bit larger than $\zeta_X$.  Three
years later, Tielens (2005) adopts $\zeta_{CR,Tiel} = 2 \times 10^{-16}$
sec$^{-1}$ per H-atom, which is $11 \, \zeta_{CR,WMHT}$!  The difference
is at least partly based on the measured abundance of the H$_3^+$ ion
(McCall et al 2002; Le Petit, Roueff, \& Herbst 2004).

We have studied a cold ($T = 17$ K) cloud that (surprisingly) lies
within the Local Bubble (Meyer, this meeting and Meyer et al.\ 2006).  I
had hoped to be able to use the observed 17 K temperature of this cloud
to establish an interesting upper limit on $\zeta_{CR}$ within the Local
Bubble. However, the equilibrium temperature varies slowly with
$\zeta_{CR}$: For $\delta_c= 1$, $T$ rises to only 14.7 K for
$\zeta_{CR}=\zeta_{CR,Tiel}$. Getting $T$ as high as 17 K requires
$\zeta_{CR} \sim 6 \zeta_{CR,Tiel}$. Evidently, the sensitivity of $T$ to
$\zeta_{CR}$ is small!  Just a little PAH heating could accomplish the
same thing.

\subsection{Optical Lines from Minority Ionization Species}

	Common optical interstellar absorption lines, like those of
NaI, are produced by minority ionization species (ionization state $r$). 
If the next-higher ionization state ($r+1$) is the dominant one, then
we have 

\begin{equation}
{n_{r} \over n_{r+1}} = {\alpha n_e \over \Gamma} \ ,
\end{equation}

\noindent where $\Gamma$ is the ionization rate of state $r$ and
$\alpha$ is the recombination coefficient from $r+1$ to $r$. If these
two ionization states comprise all of the element, then with $n_r \ll
n_{r+1}$ the interstellar line strength (and $n_r$) $\propto \alpha
n_e$. Typically $\alpha \propto T^{-0.6}$. If all the electrons come
from Carbon, then $n_e \propto n(HI) \propto {P \over T}$. The total $T$
dependence is quite steep, $T^{-1.6}$---even steeper than for the 21-cm
absorption line of HI [for which the absorption line strength $\propto
N(HI) T^{-1}$].

	Morton (1975) provides tables of $\Gamma$ and $\alpha_{56}$,
where the subscript implies $\alpha$ evaluated at $T=56$ K.  If one
derives $n_r$ at 15 K instead of 56 K, the values are $\left(15 \over 56
\right)^{-0.6} = 8.2$ times larger.  More generally, with $n_r \ll
n_{r+1}$ we have

\begin{equation}
{n_r \over  n_{tot}} \approx {\alpha_{56}\over \Gamma} \ 8.2 \ {T_{15}^{1.62} \over
\widetilde{P}_{TSAS}/\widetilde{P}_{CNM}} \ ,
\end{equation}

\noindent where $n_{tot}$ is the volume density of both ionization
states combined. For NaI and CI lines, ${\alpha_{56}\over
\Gamma} = 2.1$ and 11.7, respectively. 

All of this implies that common absorption lines from minority
ionization species in the TSAS should be strong. 

\subsection{H$_2$ Abundance}

	The H$_2$ molecule is formed on grains and destroyed by
ultraviolet photons. The photon destruction occurs in spectral lines,
which allows H$_2$ to shield itself; the destruction rate $\propto
N(H_2)^{-0.75}$. Tielens (2005) provides the numbers. For a flat slab
illuminated on one side, the local volume density H$_2$ fraction is 

\begin{equation}
{n(H_2) \over 2n(H)} = 5 \times 10^{-15} \ n(H)^4 \ N(H)_{18}^3  \ .
\end{equation}

\noindent Here $n(H)$ is the total number of H-nuclei, including both
H-atoms and H-molecules.  Integrating in from the slab's edge, the
column density fraction is 4 times smaller. 

Contrary to H97, the H$_2$ abundance is small; the difference comes from
H97's use of different numerical constants. This revision is favorable,
because now H$_2$ doesn't add to the TSAS pressure.

\subsection{Environment: Evaporation}

	TSAS clouds are enveloped by a surrounding medium. This
medium is another gas phase. Possible phases include warmer CNM, WNM,
WIM, and HIM. The interface supports either evaporation of the embedded
cloud to the environment, or condensation of the environment onto the
embedded cloud. 

	Cowie \& McKee (1977) provided the original treatment of the
spherical cloud case. Jon Slavin's talk at this meeting provides an
excellent summary, including both the essential physics and convenient
numerically-expressed equations; we use his equations in this
paper. These processes are not intuitive, as our short discussion below
illustrates. In particular, at the most basic level the mass loss rate
depends only on $\kappa$, the thermal conductivity in the external
medium, and the cloud radius $R_c$.  For ionized external media, the
effective conductivity $\kappa$ can be greatly reduced if magnetic field
lines lie parallel to the boundary. Contrary to intuition, the mass loss
rate per unit area is independent of cloud volume density $n_{cl}$. A
very important concept is heat-flux saturation, which produces an upper
limit on the evaporation rate.

For unsaturated evaporation, the most important and surprising point is
that the mass loss rate {\it per unit area} of cloud is {\it not} independent
of the radius of curvature $R_c$, but rather is proportional to
$R_c^{-1}$; this results from the scaling of the gradient for spherical
geometry. This means that, for a spherical cloud, the overall
evaporation rate $\propto R_c$ instead of $R_c^2$ and the evaporation
lifetime $\propto R_c^2$ instead of $R_c$. In contrast, the saturated
evaporation rate per unit area $\propto R_c^{1/6}$ so the lifetime
$\propto R^{7/6}$. 

A second important concept is evaporation versus condensation: small
clouds lose mass by evaporation, while large clouds gain mass by
condensation.  The critical radius $r_{crit}$ is given by Cowie \& McKee
Figure 2.  For CNM embedded in WNM at $\widetilde{P}=4000$ cm$^{-3}$ K,
$r_{crit} \sim 4000$ AU.  If TSAS clouds are spherical, then they are
smaller than this so they should evaporate.  Similarly, TSAS should also
evaporate when embedded in HIM.

But TSAS is almost certainly not spherical! The shapes of clouds really
matter for evaporation.  Thus, flat clouds, whose faces have infinite
$R_c$, never evaporate but instead grow by condensation---on their
faces. However, their edges can be sharp with small radii, where they
can evaporate. Evaporation at the edges and condensation on the faces
tends to push them towards sphericity. 

TSAS evaporation is unsaturated when embedded in WNM and saturated in
HIM. The evaporation timescale for spherical clouds having radius equal
to the canonical $L_\perp = 30$ AU of equation \ref{tsaslperp} is about
500 yr (this time $\propto R_c^2$) and 5000 yr (this time $\propto
R_c^{7/6}$) in the WNM and HIM, respectively. These are short
timescales, comparable to the dynamical timescale we estimated in \S
\ref{dynamics} Both times are
much shorter than times for thermal equilibrium.

There might not be a sharp interface. For example, TSAS clouds are
probably very cold and have no grain heating. They might be surrounded
by a thicker CNM envelope that contains grains, in which the temperature
gradually increases outwards to perhaps 100 K or more; this would be an
effective blanket against TSAS evaporation. 

\section{Summary}

Section \ref{one} provides canonical values for TSAS, assuming that it
consists of discrete structures.  Section \ref{two} compares equilibrium
versus nonequilibrium. In {\it thermal} equilibrium, with rough pressure
equality the TSAS is so cold that the cooling rates are small; the
thermal equilibrium time scale is $\sim 10^5$ yr, which might be too
long for thermal equilibrium to be established. If the TSAS is not in
thermal equilibrium, then it is probably warmer than 15 K, which
exacerbates the TSAS overpressure problem and leads to {\it dynamical
non}equilibrium. If the TSAS is not in {\it dynamical} equilibrium,
meaning that its internal pressure exceeds that of its surroundings,
then the overpressure makes TSAS clouds expand, probably forcibly. TSAS
exhibits time variability, which is not inconsistent with such
expansion. However, if the TSAS is far from equilibrium, one must
concoct a mechanism for its formation.

Section \ref{three} discusses some physical processes in TSAS. TSAS is
probably very cold, near the lower limit of $\sim 15$ K for optically thin
molecule-free gas. Optical lines of minority species get stronger at low
temperatures, and these lines should be strong in the TSAS. Evaporation
and/or dynamical expansion limits TSAS lifetimes for spherical clouds,
but maybe not for anisotropic clouds, magnetic fields, or unsharp cloud
boundaries.  Molecular hydrogen should not be plentiful in TSAS.

\acknowledgements It is a pleasure to thank Al Glassgold for helpful
discussion. This research was supported in part by NSF grant
AST-0406987.

\end{document}